\documentclass{ws-procs9x6}

\begin{document}

\title{Two aspects of Color Superconductivity: Gauge independence and Neutrality}

\author{A. GERHOLD\footnote{\uppercase{W}ork
supported by the \uppercase{A}ustrian \uppercase{S}cience \uppercase{F}oundation \uppercase{FWF}, 
project no. 16387-\uppercase{N}08.}}

\address{Institut f\"ur Theoretische Physik, Technische Universit\"at Wien,\\
Wiedner Hauptstr. 8-10, A-1040 Vienna, Austria\\ E-mail: gerhold@hep.itp.tuwien.ac.at}

\maketitle

\abstracts{A formal proof is given that the fermionic quasiparticle dispersion laws in a color
superconductor are gauge independent. It is shown that the gluon (photon) field acquires a
non-vanishing expectation value in a color superconductor, which is related to color
(electric) neutrality.}

\section{Introduction}
It is well known that cold dense quark matter is a color superconductor.\cite{Bailin:1984bm} 
Color superconducting
phases are characterized by a non-vanishing gap function, which can be calculated self-consistently
from the Schwinger-Dyson equation (gap equation). At asymptotic densities, calculations from
first principles (QCD) should be possible.
However, calculations of the gap function seem to give different results
for covariant and Coulomb gauge.\cite{Pisarski:1999tv} This was the motivation to construct a 
formal proof of gauge independence,
which is presented in Sec. 2. It turns out that this proof involves non-vanishing expectation 
values of the gauge
fields. These expectation values act as effective chemical potentials for
color and electric charge, which will be briefly discussed in Secs. 3-5.


\section{Gauge independence of fermionic quasiparticle dispersion laws}
The inverse quark propagator in the Nambu-Gor'kov basis is given by
\begin{eqnarray}
  &&\mathcal S^{-1}=\left(
    \begin{array}{cc} Q\!\!\!\!/+\mu\gamma_0
    +\Sigma & \Phi^- \\
    \Phi^+ & Q\!\!\!\!/-\mu\gamma_0+\bar\Sigma 
    \end{array}
  \right)\nonumber\\
  &&\qquad=\int d^4x\, e^{-iQ\cdot x}
  \frac{\delta^2\Gamma}{\delta\bar\Psi(x)\delta\Psi(0)}\Big|_{\psi=\bar\psi=A_i^a=0,
  A_0^a=\tilde A_0^a},
\end{eqnarray}
where $\Phi^+$ is the gap function, $\Phi^-(Q)=\gamma_0 [\Phi^+(Q)]^\dag\gamma_0$, 
$\Sigma$ is the quark self energy, $\bar\Sigma(Q)=C[\Sigma(-Q)]^T C^{-1}$ with the charge 
conjugation matrix $C$,
$\Gamma$ is the effective action, and $\tilde A$ is the expectation value of the gluon field.
The change in the effective action under the variation of some gauge parameter(s)
is given by the following gauge
dependence identity,\cite{Nielsen:1975fs,KKR}
\begin{equation}  
  \delta\Gamma= \Gamma_{,i}\delta X^i+\int dx\frac{\delta\Gamma}{\delta A^{a\mu}(x)}
  \delta X_{(A)}^{a\mu}(x). \label{gl2}
\end{equation}
We use the DeWitt notation for the fermions, $i=(\psi(x), \psi_c(x))^T$, 
$\bar i=(\bar\psi(x), \bar\psi_c(x))$.  The $\delta X$'s in Eq. (\ref{gl2}) are 
quite complicated functionals,
whose diagrammatic expansion is shown in Ref.~\refcite{KKR}. 
After taking the second derivative of Eq. (\ref{gl2}), switching to momentum 
space,\footnote{Here we assume translation invariance, therefore our proof is not directly applicable for
inhomogeneous color superconducting phases, which might appear at intermediate densities, see e.g. 
Ref. \refcite{Casalbuoni:2003wh}.}
and taking the determinant with respect to Nambu-Gor'kov, Dirac, color and flavor indices, 
we obtain\cite{Gerhold:2003js} 
\begin{equation}
  \delta\det(\Gamma_{i\bar j})+\delta\tilde A^{a0}
            \frac{\partial}{\partial \tilde A^{a0}}\det(\Gamma_{i\bar j})
            =-\det(\Gamma_{i\bar j})[\delta X^k_{\,,k}
            +\delta X^{\bar k}_{\,,\bar k}].
\end{equation}
This identity means that the total variation of the determinant of the inverse propagator is proportional
to the determinant of the inverse propagator itself. 
Therefore the fermionic quasiparticle dispersion laws are gauge independent,
provided that the singularities of $\delta X$ do not coincide with those of the quark propagator.
In principle $\delta X$ could develop mass shell singularities, as it is the case for the
fermion damping rate in high temperature QCD.\cite{Baier:1992dy}
Then a gauge independent result could only be obtained by introducing an IR cutoff, to be lifted
only at the very end of the calculation.\cite{Rebhan:1992ak}
However, in the case of color superconductivity a simple IR cutoff turns out not to be sufficient
to remove the gauge dependence of the gap
in covariant gauges.
Indeed, it has been argued that this gauge dependence will
only disappear after the inclusion of vertex corrections in the gap equation.\cite{Hou:2004bn}

\section{Gluon tadpole and color neutrality}
In contrast to NJL models, where color neutrality has to be imposed as an external 
condition,\cite{Amore:2001uf}
the color superconducting phases in full QCD are automatically color neutral. This can be demonstrated in 
the following way.\cite{Gerhold:2003js,Khlebnikov:1996vj}
The fields $A_0^a$ appear in the QCD action as Lagrange multipliers for the Gauss law constraint
(at least in a gauge which does not involve $A_0^a$.)
 Therefore in the path integral the integration over the zero-momentum modes $A_
{0,\vec p=0}^a$  
produces delta functions, $\delta(N_a)$, where $N_a$ are the color charges. This
 means that
only color neutral field configurations contribute to the partition function.

In a color superconductor the gluon field acquires a non-vanishing expectation 
value.\cite{Gerhold:2003js,Kryjevski:2003cu,Dietrich:2003nu}
The leading order contribution to this expectation value can be computed from the one-loop gluon
tadpole diagram. For the 2SC phase we find\cite{Gerhold:2003js}
\begin{equation}
  \mathcal{T}^a\simeq-\delta^{a8}\frac{2g}{\sqrt3\pi^2}\, \mu\,\phi^2
  \,\mathrm{ln}\left(\frac{\phi}{2\mu}\right). \label{gl6}
\end{equation}
The expectation value of $A_0^a$ can be obtained by attaching the external gluon propagator, which
gives\cite{Gerhold:2003js,Dietrich:2003nu}
\begin{equation}
  \tilde A_0^8\sim\phi^2/(g^2\mu).
\end{equation}
This expectation value acts as an effective chemical potential for the color charge 
(via $\mu_8=gA_0^8$), thus ensuring color neutrality.\cite{Dietrich:2003nu}

In Eq. (\ref{gl6}) we have computed the quantity
\begin{equation}
  \mathcal {T}^a\sim\frac{\delta\Gamma}{\delta A_0^a}\bigg|_{A=0}.
\end{equation}
We remark that at the order of our computation
this calculation is completely equivalent to the self-consistent approach
of Ref. \refcite{Dietrich:2003nu}, where $\tilde A$ has been computed from the Yang-Mills equation
\begin{equation}
  \frac{\delta\Gamma}{\delta A_0^a}\bigg|_{A=\tilde A}=0.
\end{equation}
The connection between the two approaches can be seen from the Taylor expansion
\begin{equation}
  \frac{\delta\Gamma}{\delta A_0^a}\bigg|_{A=0}=
  \frac{\delta\Gamma}{\delta A_0^a}\bigg|_{A=\tilde A}-
  \frac{\delta^2\Gamma}{\delta A_0^a\delta A_0^b}\bigg|_{A
  =\tilde A}\tilde A_0^b
  +\ldots
\end{equation}
On the right hand side the first term vanishes, while the second term is essentially
the inverse gluon propagator at zero momentum times the expectation value of $A$.

\section{Gluon tadpole for CFL with $m_s\neq0$}

In the CFL phase one finds $\tilde A_0^a=0$ at one-loop order for $m_s=0$.
For finite $m_s$ the quark propagator in the Nambu-Gor'kov basis is given by
\begin{equation}
  \mathcal{S}=\left(\begin{array}{cc}[G^-_0]^{-1}
    & \Phi^- \\
    \Phi^+ & [G^+_0]^{-1}
    \end{array}
  \right)^{-1}
\end{equation}
with
\begin{equation}
  [(G^\pm_0)^{-1}]_{rs}^{ij}=[(Q\!\!\!\!/\pm\mu\gamma_0)(\delta_{rs}-\delta_{r3}\delta_{s3})+
  (Q\!\!\!\!/\pm\mu\gamma_0-m_s)\delta_{r3}\delta_{s3}]\delta^{ij}, \
\end{equation}
where the lower indices are flavor indices and the upper indices are color indices.
At finite $m_s$ the Dirac structure of the gap is more complicated than in the
massless case,\cite{Pisarski:1999av} but for $m_s\ll\mu$
we may assume that the additional terms are suppressed 
at least with $m_s/\mu$,\cite{Fugleberg:2002rk}
\begin{equation}
  [\Phi^+]_{rs}^{ij}\simeq\phi_+\Lambda^+_{\bf q}\gamma_5
  (\delta^i_r\delta^j_s-\delta^j_r\delta^i_s)+\mathcal{O}\left(\frac{m_s}{\mu}\right).
\end{equation} 
Assuming $m_s\gg\phi_+$ we find for the tadpole diagram at leading order in $m_s$
\begin{eqnarray}
  &&\!\!\!\!\!\!\!\!\!\!\!\mathcal{T}^a\simeq\delta^{a8}
  \frac{gm_s^2}{2\sqrt{3}\pi^3}\int_{-\infty}^\infty dq_4\int_0^\infty dq\,q\,\phi_+^2\nonumber\\
  &&\qquad\times\frac{q_4^2-3(q-\mu)^2-8\phi_+^2}
  {\left(q_4^2+(q-\mu)^2+\phi_+^2\right)^2
  \left(q_4^2+(q-\mu)^2+4\phi_+^2\right)}\nonumber\\
  &&\simeq\delta^{a8}\frac{g\mu m_s^2}{18\sqrt{3}\pi^2}(-21+8\ln2),
\end{eqnarray}
where we have neglected the momentum dependence of the gap function.
Attaching the gluon propagator with the Debye mass\cite{Rischke:2000ra} proportional to $g\mu$ we find
as in Ref. \refcite{Kryjevski:2003cu}
\begin{equation}
  \tilde A_0^a\sim \delta^{a8}\frac{m_s^2}{g\mu}.
\end{equation}
Therefore the effective chemical potential is
$\mu_8\sim m_s^2/\mu$, which is consistent with results from NJL model 
calculations.\cite{Amore:2001uf}

\section{Photon tadpole and electric neutrality}
If one takes into account also electrons and photons, electric neutrality will we achieved
by a mechanism which is completely analogous to the one described in Sec. 3.
For the expectation value of the photon field one finds in the 2SC phase (for $m_s=0$)
\begin{equation}
  A^0\sim \frac{\phi^2}{e\mu}\ln\left(\frac{\phi}{2\mu}\right).
\end{equation}

\section*{Acknowledgements}
I would like to thank C. Manuel and A. Rebhan for very helpful discussions.

\end{document}